\documentclass[%
 reprint,
 superscriptaddress,
 amsmath,amssymb,
 aps,
 prb,
 floatfix,
]{revtex4-2}

\usepackage{graphicx}
\usepackage{dcolumn}
\usepackage{bm}
\usepackage{tikz}
\usepackage{gensymb}
\usepackage{upgreek}
\usepackage{amsmath, amsfonts, amssymb}  
\usepackage{siunitx}  

\begin{document}


\title{Experimental Generation of Optimally Chiral Azimuthally-Radially Polarized Beams}

\author{Albert Herrero-Parareda}
\affiliation{Department of Electrical Engineering and Computer Science, University of California, Irvine, CA, USA}

\author{Nicolas Perez}
\affiliation{Beckman Laser Institute, University of California, Irvine, CA, USA}

\author{Filippo Capolino}
\affiliation{Department of Electrical Engineering and Computer Science, University of California, Irvine, CA, USA}

\author{Daryl Preece}
\affiliation{Beckman Laser Institute, University of California, Irvine, CA, USA}
\email{dpreece@uci.edu}

\date{\today}

\begin{abstract}
We implement a paraxial azimuthally-radially polarized beam (ARPB), a novel class of structured light beams that can be optimal chiral (OC), leading to maximum chirality density at a given energy density. By using vectorial light shaping techniques, we successfully generated a paraxial ARPB with precise control over its features, validating theoretical predictions. Our findings demonstrate the ability to finely adjust the chirality density of the ARPB across its entire range by manipulating a single beam parameter. Although our experimental investigations are primarily focused on the transverse plane, we show that fields whose transverse components satisfy the optimal chirality condition are optimally chiral in all directions, and our results highlight the promising potential of OC structured light for applications in the sensing and manipulation of chiral particles. We show that optical chirality is more general than the concept of handedness. This work represents a significant advancement toward practical optical enantioseparation and enantiomer detection at the nanoscale.
\end{abstract}


\maketitle

\section{Introduction} 
\label{ch:Intro}

Part of this study was inspired by the work of Prof. Federico Capasso, to whom this special issue is dedicated. The many topics Prof. Capasso worked on include helicity, chirality of light and the interaction with chiral matter, most notably in Refs.~\cite{hayat_lateral_2015, zhu_giant_2018, lu_helicity_2023}. Some of the material of this paper was presented and discussed during the 2024 NanoPlasm Conference in Cetraro (IT) where Prof. Capasso's $75$th birthday was celebrated. 

The study of structured light is important for various applications, including the detection of chiral nanoparticles \cite{forbes_enantioselective_2022, rosales_light_2012, forbes_customized_2023}. Chirality is a property of objects that are not superimposable with their mirror image \cite{barron_true_2013}. Importantly, many biologically relevant molecules exist in chiral pairs, known as enantiomers \cite{lininger_chirality_2022}. The chiral nature of electromagnetic fields can be described using the optical chirality density \cite{tang_optical_2010, bliokh_characterizing_2011, forbes_customized_2023}, or the helicity density \cite{alpeggiani_electromagnetic_2018, poulikakos_optical_2019, forbes_optical_2022}. 
Traditionally, the "handedness" of circularly polarized light has been associated with the chirality of light, but here we show that the chirality density of light is much more general than the simple concept of handedness. We employ the concept of the {\it helicity density} because it is easily associated with physical quantities of electromagnetic fields \cite{trueba_electromagnetic_1996, mackinnon_on_2019}. Specifically, in quantum mechanics, the helicity of a light field is characterized as the projection of the spin angular momentum onto the direction of propagation \cite{alpeggiani_electromagnetic_2018}. For monochromatic beams with the implicit time dependence $e^{-i\omega t}$, where $\omega$ is the angular frequency of light, the time-average helicity density $h$ is \cite{cameron_optical_2012, hanifeh_optimally_2020, fang_optical_2021}

\begin{equation}
    h = \frac{1}{2\omega c}\Im{\left(\mathbf{E}\cdot\mathbf{H}^*\right)},
    \label{eq:HelicityDensityMonochromatic}
\end{equation}

where $\mathbf{E}$ and $\mathbf{H}$ represent the electric and magnetic field phasors of light, respectively. The term $c=1/\sqrt{\varepsilon_0\mu_0}$ is the speed of light in vacuum. The time-average chirality density $C$ is proportional to the time-average helicity density $h$ for monochromatic beams, i.e., $C=\omega^2 c^{-1} h$ \cite{hanifeh_optimally_2020}.

In Ref.~\cite{hanifeh_optimally_2020}, it was shown that the magnitude of the helicity density of a monochromatic field, at a given time-average energy density $u=\varepsilon_0|\mathbf{E}|^2/4 + \mu_0|\mathbf{H}|^2/4$ \cite{angelsky_structured_2020}, has an upper bound, i.e., $|h|\le u/\omega$ always. Light fields that reach the upper bound $|h|= u/\omega$ are known as optimal chiral light (OCL). Circularly polarized light (CPL) is the most intuitive example of OCL \cite{lininger_chirality_2022}, and the sign of $h$ is related to the handedness of the CPL. The same upper bound for the magnitude of the helicity density was stated in Ref.~\cite{bliokh_dual_2013} involving fields whose Fourier spectrum representation contains only plane waves with one circular polarization. However, the concepts of helicity density and optimal chirality hold true for any kind of monochromatic structured light, including cases where the magnetic and electric fields are polarized along a single (e.g., the beam's longitudinal) direction, and the concept of handedness cannot be applied. The necessary and sufficient condition for fields to be locally optimally chiral is

\begin{equation}
    \mathbf{E}=\pm i\eta_0\mathbf{H},
    \label{eq:OptimalChiralityCondition}
\end{equation}

where $\eta_0 =\sqrt{\mu_0/\varepsilon_0}$ is the intrinsic impedance of free space \cite{hanifeh_optimally_2020}. This condition, referred to as the optimal chirality condition, stipulates that optimally chiral fields are those whose electric and magnetic field phasors have balanced magnitudes and a quarter-period phase delay between them. Fields that satisfy this condition also display a remarkable electric-magnetic symmetry in their energy and spin densities \cite{hanifeh_optimally_2020}. Under this optimal chirality condition, one has $h=\pm \mu_0 |\mathbf{H}|^2/(2\omega)=\pm \varepsilon_0 |\mathbf{E}|^2/(2\omega)$. 

The concepts of optimal chirality and self-duality are equivalent for monochromatic beams. Self-duality refers to fields that are unchanged by the the duality transformation $\mathbf{E}\to \mathbf{B}$ and $\mathbf{B}\to -\mathbf{E}$ \cite{bliokh_dual_2013, fernande_intro_2015}. These fields are eigenvectors of the curl \cite{lekner_selfdual_2019}, i.e., $\nabla\times \mathbf{E}=k\mathbf{E}$. As shown in Ref.~\cite{parareda_arpb_2024} without connecting the concepts of optimal chirality and self-duality, monochromatic fields satisfying the optimal chirality condition from Eq.~(\ref{eq:OptimalChiralityCondition}) are eigenvectors of the curl operator, and therefore self-dual fields. Here we use the concept of optimal chirality instead of self-duality because we focus on the chirality features of the beam, rather than on the broader electromagnetic symmetries displayed by self-dual beams. Additionally, few self-dual fields seem to have been studied experimentally \cite{lekner_self_dual_2024}.

Optimally chiral structured beams are important because they combine two powerful effects: vectorially shaped light and maximized chirality density at a given energy density. This combination is advantageous because it allows for the control of enhanced interaction of the beam with chiral matter. As a result, optimally chiral structured beams open new possibilities for controlled sensing and manipulation of chiral particles. Moreover, the topology of tailored beams enables creative designs for chirality-discriminating optical traps \cite{parareda_optimally_2023}, which aim at trapping an enantiomer while repealing its mirror image \cite{tkachenko_helicity_2014, bradshaw_chirality_2015, zhang_alloptical_2017, li_enantioselective_2019, fang_optical_2021, forbes_enantioselective_2022}. 

The unprecedented control over the amplitude and phase of structured light \cite{angelsky_structured_2020, perez_enhanced_2024} also results in an exceptional ability to finely tune the helicity density $h$ of a probing beam, shown in Eq.~(\ref{eq:HelicityDensityMonochromatic}). This precise control is crucial for the sensing capabilities of the system since it allows one to tune the interaction between the chiral particle and the field by adjusting $h$ \cite{lininger_chirality_2022}. For example, this ability leads to a more detailed characterization of the interactions between the chiral sample and the fields, even beyond the commonly used dipolar approximation. The importance of higher-order multipoles on chiral interactions is investigated in Ref.~\cite{mun_importance_2019}. Additionally, this control over the helicity density enables rapid changes in its sign (similar to reversing the "handedness" of circularly polarized light), facilitating the creation of dynamic optical potentials for enantioseparation and the experimental investigation of the chiral effects of higher-order multipoles. Dynamic optical traps \cite{bennett_spatially_2013} that enable real-time interaction with achiral particles have been designed based on holograms \cite{pleguezuelos_holotrap_2007} and diffractive elements \cite{cojoc_dynamic_2004}, while the chirality-discriminating photoinduced forces on particles under the dipolar approximation are described in Ref.~\cite{hayat_lateral_2015}.

\begin{figure}
    \centering
    \includegraphics[width = 0.45\textwidth]{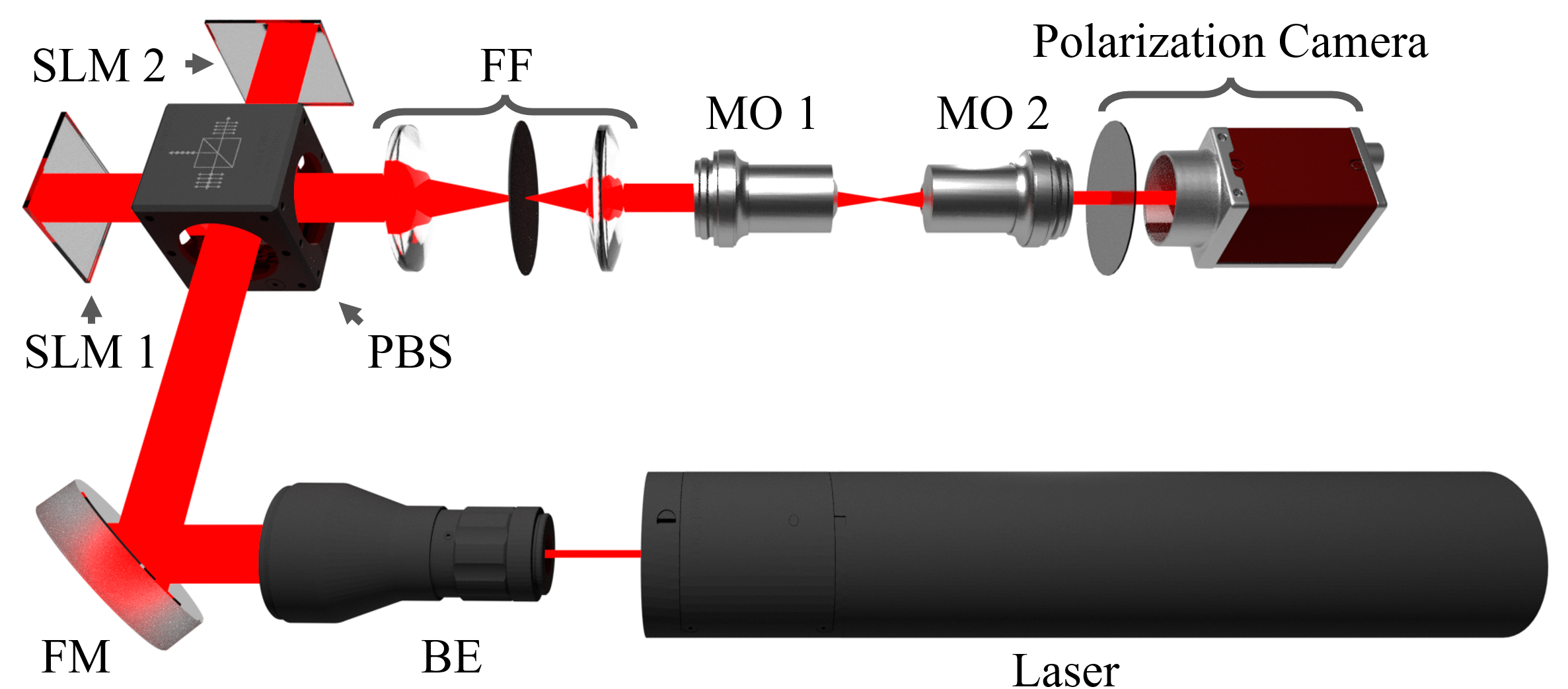}
    \caption{Diagram illustrating the experimental setup used to generate the ARPBs with variable $\psi$. A diagonally polarized He-Ne laser is first collimated and expanded by a beam expander (BE), then aligned by folding mirrors (FM) before vertical and horizontal polarizations are separated by a polarized beam splitter (PBS) onto two twin spatial light modulators (SLM1 and SLM2). Each SLM dynamically modulates the power and phase of the beam returning through the PBS. Separate calculated phases are applied to the orthogonally polarized beams before recombination at the PBS. The beam is then Fourier filtered (FF), and focused using a 0.25NA objective (MO 1). Polarization imaging is performed with a 0.85NA objective (MO 2) in the focal plane of the ARPB, which is captured using a Kiralux Polarization Camera. Updating the ARPB with a new value of $\psi$ simply involves displaying holograms generated with the modified transverse electric field components on the SLMs.
    }
    \label{fig:opticalsetup}
\end{figure}

However, the ability to generate optimally chiral structured beams is constrained. One must design an optical beam that satisfies the optimal chirality condition from Eq.~(\ref{eq:OptimalChiralityCondition}) and effectively implement it. While for a simple Gaussian beam the trivial choice would be that of circular polarization, the choice of the polarization and topology of vector beams to achieve optimal chirality is not entirely straightforward. For that reason, we have chosen to implement a previously proposed example of a structured beam that displays optimal chirality: the azimuthally-radially polarized beam (ARPB) \cite{kamandi_unscrambling_2018, hanifeh_helicity_max_planar_2020, hanifeh_optimally_2020}. The ARPB consists of a phase-shifted combination of an azimuthally polarized beam and a radially polarized beam. It has been theoretically studied in the past, see Refs.~\cite{kamandi_unscrambling_2018, hanifeh_optimally_2020, hanifeh_helicity_max_2020, jiang_theory_2021}, and most comprehensively in Ref.~\cite{parareda_arpb_2024}. The optimally chiral ARPB (OC-ARPB) combines the extraordinary properties of OCL with the spatial separation between its transverse fields, which vanish on the beam axis, and the longitudinal fields. OCL is present along the beam axis, solely due to $E_z$ and $H_z$. Consequently, the ARPB has the potential to be used for controlled, on-axis separation of enantiomers or for enantiomer detection as envisioned in Refs.~\cite{kamandi_unscrambling_2018, hanifeh_optimally_2020, hanifeh_helicity_max_2020, jiang_theory_2021}. The ARPB has also been recently studied in Refs.~\cite{eisman_exciting_2018} and ~\cite{ram_probing_2024}. In the former, the authors tightly focus an achiral combination of an azimuthally and a radially polarized beams to excite a chiral dipole moment in a geometrically achiral nanostructure, and in the latter, the authors use the ARPB to generate transverse spin angular momentum in optical tweezers. While these studies implemented the ARPB, they do not focus on analyzing the chiral features of the ARPB.

Our work presents an experimental implementation of the ARPB, with a focus on characterizing its chirality and demonstrating its ability to achieve optimal chirality \cite{hanifeh_optimally_2020,parareda_arpb_2024}. By employing advanced vectorial shaping techniques, we have overcome limitations to the field's stability and local polarization control to successfully generate a paraxial APRB with precise manipulation over its features, validating theoretical predictions. Importantly, we show that the helicity density of the ARPB can be tuned across its full range of possible values by varying a single beam parameter. While our findings are confined to studying the chirality density in the transverse plane, they demonstrate the potential of optimally chiral structured light for designing enantioseparating optical traps and advancing practical schemes for the sensing and manipulation of chiral particles. Additionally, we show in Section \ref{ch:TransverseOC} that if the transverse fields of a beam satisfy the optimal chirality condition from Eq.~(\ref{eq:OptimalChiralityCondition}), the longitudinal fields satisfy it as well.

\section{Methods}
\label{ch:Methods}
\subsection{Helicity Density}
The field phasors of the ARPB are \cite{parareda_arpb_2024}

\begin{widetext}
\begin{equation}
\begin{array}{c}
        \mathbf{E} = \frac{fV}{kw^2}\left[k\rho \left(A_\rho + iB_\rho\right)\,\hat{\bm{\rho}} + k\rho \hat{V}e^{i\psi}\,\hat{\bm{\varphi}}+2i\left(A_z + iB_z\right)\,\hat{\bm{z}}\right], \\
        \mathbf{H} = - \frac{fV}{kw^2\eta_0}\left[k\rho \hat{V}e^{i\psi}\left(A_\rho + iB_\rho\right)\,\hat{\bm{\rho}} - k\rho\,\hat{\bm{\varphi}}+2i\hat{V}e^{i\psi}\left(A_z + iB_z\right)\,\hat{\bm{z}}\right],
\end{array}
\label{eq:ARPBDef}
\end{equation}
\end{widetext}

where $V$ is a complex amplitude with units of Volts. The parameters $\hat{V}$ and $\psi$ represent the relative amplitude and phase between the electric and magnetic azimuthal components, respectively, normalized by the characteristic impedance $\eta_0$. This relationship is expressed as $E_\varphi / (\eta_0 H_\varphi) = \hat{V}e^{i\psi}$. The dimensionless shorthand parameters $f, A_\rho, B_\rho, A_z,$ and $B_z$ are

\begin{equation}
\begin{array}{c}
    f = \frac{2}{\sqrt\pi}e^{-(\rho/w)^{2}\zeta}e^{-2i\tan^{-1}(z/z_{\text{R}})}e^{ikz}, \\
    A_\rho = 1+\frac{1}{kz_{R}}\frac{\rho^{2}-2w_{0}^{2}}{w^{2}} +\left(\frac{2z\rho}{w^2 k z_R }\right)^2, \\
    B_\rho = -\frac{4}{(kw)^{2}}\frac{z}{z_{R}}\left(1-\frac{\rho^{2}}{w^{2}}\right), \\
    A_z = 1-\frac{\rho^2}{w^2}, \\
    B_z = \frac{z}{z_R}\frac{\rho^2}{w^2},
\end{array}
\label{eq:Simplification}
\end{equation}

where $w$ is the beam radius, defined as $w=w_0\sqrt{1+(z/z_R)^2}$, and $w_0$ is defined as half the beam waist parameter at $z=0$. The Gouy phase is $\zeta = 1-iz/z_R$, and the Rayleigh range is denoted as $z_R = \pi w_0^2/\lambda$, where $\lambda$ is the wavelength in free space. The wavenumber is $k=2\pi/\lambda$. While $f$ is a complex scalar, the other parameters $A_\rho$, $B_\rho$, $A_z$, and $B_z$ in Eq.~(\ref{eq:Simplification}) are real valued.

Adjusting the phase-shift $\psi$ (referred to as the phase parameter of the ARPB) and the relative amplitude $\hat{V}$ enables the creation of an ARPB that meets the optimal chirality condition described in Eq.~(\ref{eq:OptimalChiralityCondition}). This specific configuration occurs for $\psi = \pm\pi/2$ and $\hat{V}=1$ \cite{hanifeh_optimally_2020, parareda_arpb_2024}. Note that the OC-ARPB is a structured, monochromatic, and self-dual beam.

Most notably, the ARPB has transverse fields that vanish on the beam axis ($\rho=0$), where the longitudinal fields persist \cite{kamandi_unscrambling_2018, hanifeh_optimally_2020, hanifeh_helicity_max_2020, jiang_theory_2021}. This spatial separation between the transverse and longitudinal components of the beam results in vanishing linear and angular momentum densities on the beam axis \cite{parareda_arpb_2024}, where only the energy and helicity densities ($u$ and $h$ respectively) associated with the longitudinal fields persist. For the ARPB, the time-average energy and helicity densities across the entire beam are \cite{parareda_arpb_2024}

\begin{equation}
    \begin{array}{c}
          u = u_0 D (1 + \hat{V}^2)/2 , \\
          h = h_0 D  \hat{V} \sin \psi, 
    \end{array}
    \label{eq:EnergyHelicityARPB}
\end{equation}

where

\begin{equation}
             D = (k\rho)^2\left(1 + A_\rho^2 + B_\rho^2\right) + 4\left(A_z^2 + B_z^2\right),
    \label{eq:Parameters}
\end{equation} 

and $u_0 = \frac{\varepsilon_0}{2k^2w^4}|f|^2|V|^2$ and $h_0 = \frac{\varepsilon_0}{2\omega k^2w^4}|f|^2|V|^2$ are normalization constants with units of energy density ($\text{J}/\text{m}^3$) and helicity density ($\text{Ns}/\text{m}^2$), respectively. They are related as $h_0=u_0/\omega$, leading to 

\begin{equation}
    h=\frac{u}{\omega} \frac{2\hat{V}}{1 + \hat{V}^2}  \sin \psi.
    \label{eq:hurelation}
\end{equation} 

\begin{figure*}[htbp]
    \centering
    \newcolumntype{P}[1]{>{\centering\arraybackslash}p{#1}}
    \setlength\tabcolsep{0pt} 
    \begin{tabular}{*{6}{P{0.18\textwidth}}}
        \includegraphics[width=0.18\textwidth]{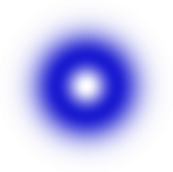} &
        \includegraphics[width=0.18\textwidth]{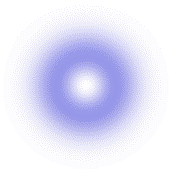} &
        \includegraphics[width=0.18\textwidth]{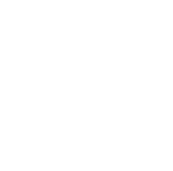} &
        \includegraphics[width=0.18\textwidth]{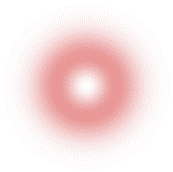} &
        \includegraphics[width=0.18\textwidth]{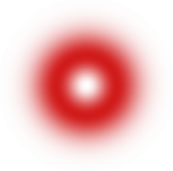} &
        \includegraphics[width=0.075\textwidth]{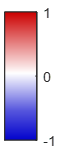}\\
        
        (a) $\psi=-\pi/2$ & (b) $\psi=-\pi/4$ & (c) $\psi=0$& (d) $\psi=\pi/4$& (e) $\psi=\pi/2$ \\
        \includegraphics[width=0.18\textwidth]{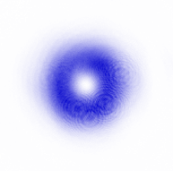} &
        \includegraphics[width=0.18\textwidth]{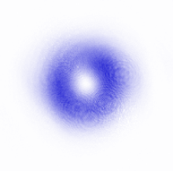} &
        \includegraphics[width=0.18\textwidth]{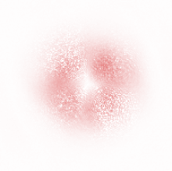} &
        \includegraphics[width=0.18\textwidth]{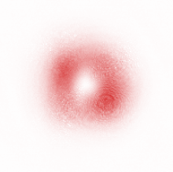} &
        \includegraphics[width=0.18\textwidth]{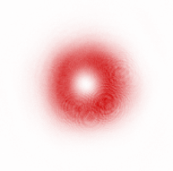} &
        \includegraphics[width=0.075\textwidth]{Pictures/colorbar_v2.png} \\
        (f) $\psi=-\pi/2$ & (g) $\psi=-\pi/4$ & (h) $\psi=0$& (i) $\psi=\pi/4$& (j) $\psi=\pi/2$ \\
    \end{tabular}
    \caption{Predicted (above) and experimental (below) third normalized Stokes parameter $s_3$ on the transverse plane for ARPBs with $\psi=-\pi/2, -\pi/4, 0, \pi/4, \pi/2$ (and unity relative amplitude $\hat{V}=1$). The first row, (a)-(e), displays theoretical $S_3$ at focus, while the second row, (f)-(j), presents the experimental results. As indicated by the colorbars on the right side of the figure, blue denotes a negative $s_3$, while red indicates a positive $s_3$. For the experimental results, the sign of $s_3=\pm|s_3|$ has been chosen to better visually represent the change in the $s_3$ of an ARPB with $\psi$.}
    \label{fig:ThirdStokesParameter}
\end{figure*}

\subsection{Vectorial Beam Shaping}
The paraxial ARPB is produced experimentally using spatial light modulators (SLMs), which introduce a digitally controlled spatially variable phase shift $\phi(x,y)$ to an incident optical field. The pixelated liquid crystal cells in an SLM differentially delay the phase of incident fields according to the voltage applied over each pixel. The SLM is controlled via a standard computer. For a calculated scalar field $E$, the hologram is produced by converting calculated grayscale values into calibrated phase values on the SLM screen.

Although amplitude modulation is not directly available with a phase-only SLM, pseudo-amplitude modulation is possible. Amplitude modulation can be created using a scattering mask; the inverse magnitude of the field $1-|E|/\textrm{max}|E|$ is multiplied by random integers and then multiplied onto the grayscale phase image, scattering unwanted power into higher diffraction orders (see Ref.~\cite{stilgoe_interpretation_2016} for details). Lastly, a blazed grating is applied to tilt the shaped beam and preferentially directs power into the first-order diffraction spot \cite{Liesener_multi_2000}. The first order is then spatially filtered to eliminate non-diffracted light (in the zero order) and unwanted higher harmonics.

Most commonly, SLMs are used to produce scalar waves with a constant polarization direction. Here, however, we use two SLMs to produce a vectorially shaped beam. The paraxial ARPB is generated using a twin SLM setup illustrated in Fig.~\ref{fig:opticalsetup}, initially introduced in Ref.~\cite{perez_enhanced_2024} to generate obscured bottle beams. This method uses the twin SLMs to holographically control the beam's transverse field components, $E_x$ and $E_y$, separately.

To create such vectorially shaped beams, we first obtain the complex field components of the associated Jones vector $\left[  {E}_{x},  {E}_{y} \right]^T=\left[ |E_{x}|e^{i\phi _{x}}, |E_{y}| e^{i\phi _{y}} \right]^T$, where $T$ denotes the transpose operation, and separately convert each of them into holograms. These are displayed on two corresponding orthogonally aligned SLMs, which are aligned to match the polarization of the field component they display. The SLMs in this configuration provide control of phase and intensity of $E_x$ and $E_y$ individually, resulting in control of the local polarization state of the structured beam. The difference in amplitude modulation on each SLM rotates the orientation angle of the polarization according to $\arctan (|E_y|^2/|E_x|^2)$, while phase shifting $\phi _{x}$ relative to $\phi _{y}$ controls the ellipticity of the polarization state. The combination of these methods allows for local control of phase, amplitude, and polarization of the beam concurrently \cite{preece_independent_2008}. It is notable that to maintain system stability, a fixed optical path length must be preserved between each SLM. Although relative phase or position changes will not cause interference between the orthogonal field components, such changes can alter the outgoing polarization angle and ellipticity. Therefore, the system has been engineered to remain fixed in place.

In this paper, we produce five different paraxial ARPBs (which are assumed to have ($E_z=0$) under the zeroth-order approximation \cite{lax_maxwell_1975}), with unity relative amplitude $\hat{V}=1$ and a phase parameter $\psi$ ranging from $-\pi/2$ to $\pi/2$. Generating or updating ARPBs with a new value of $\psi$ simply involves processing holograms with the appropriate electric field components for display on the SLMs.

\subsection{Helicity and Stokes Parameters}
For each beam we simultaneously record the intensities $I$ for the horizontal $x$, vertical $y$, diagonal $d$, and anti-diagonal $a$ polarizations by using a polarization-sensitive camera (Thorlabs, Kiralux). 
These polarizations are at $0^\circ, 90^\circ, 45^\circ, -45^\circ$ with respect to the horizontal $x$ axis, respectively. 
From the polarized intensity measurements we calculate the Stokes parameters $S_0$, $S_1$, and $S_2$ \cite{mcmaster_polarization_1954,berry_measurement_1977}, i.e., 

\begin{equation}
    \begin{array}{c}
        S_0 = \left(I_x + I_y\right) = \left(I_d + I_a\right), \\
        S_1 = \left(I_x - I_y\right), \\
        S_2 =  \left(I_d - I_a\right).
    \end{array}
    \label{eq:StokesDefinition}
\end{equation}

The intensity is defined herein as the squared of the field amplitudes \cite{fatadin_accurate_2006, hui_basic_2009}, i.e., $I = |\mathbf{E}|^2$. Here we define the normalized Stokes parameters as $s_i = S_i/S_0$ for $i = 1, 2, 3$. For monochromatic beams, they are related as \cite{born_principles_1999}

\begin{equation}
    \begin{array}{c}
        s_1^2 + s_2^3 + s_3^2 = 1,
    \end{array}
    \label{eq:MonochromaticStokesRelation}
\end{equation}

where $S_3=I_{LCP}-I_{RCP}$ is the difference between the intensity of left-handed and right-handed CPL, and $s_3=S_3/S_0$. Therefore, we can extract the magnitude of the third normalized Stokes parameter $|s_3|=1- s_1^2 - s_2^3$  from linear polarization measurements. This normalized Stokes parameter is directly related to the helicity density of paraxial beams whose longitudinal fields are neglected (see Appendix~A for details)

\begin{equation}
    s_3  = h\omega/u.
    \label{eq:StokesS3andh}
\end{equation}

The concept of optimal chirality states that the magnitude of the helicity density $h$ for any kind of monochromatic structured light has the upper bound of $u/\omega$, as demonstrated in Ref.~\cite{hanifeh_helicity_max_2020}. Therefore, we find it convenient to use the concept of the normalized helicity density $\hat{h} = h\omega/u$, as in Ref.~\cite{parareda_arpb_2024}, whose value is bounded by $ -1 \leq \hat{h} \leq 1$. We conclude that under the paraxial approximation, $s_3=\hat{h}$, and that $|s_3|\leq1$ is consistent with the concept of OCL that states that for any structured light $|\hat{h}|\leq1$.

For the ARPB, the normalized helicity density is

\begin{equation}
    \hat{h} =\frac{2\hat{V}}{1+\hat{V}^2}\text{sin}\psi,
    \label{eq:ARPBNormalizedHelicityDensity}
\end{equation}

and when we use unity relative amplitude, $\hat{V}=1$, we have $\hat{h}=\text{sin}\psi$. In this work, we restrict our analysis to paraxial beams with negligible longitudinal fields, for which the normalized helicity density is equivalent to the third normalized Stokes parameters, $\hat{h}= s_3$. In Section~\ref{ch:Results}, we will experimentally verify that $h=s_3=\text{sin}\psi$ for the paraxial ARPB.
For non-paraxial beams, which have a helicity density with a contribution from the non-negligible longitudinal fields, $\hat{h} \neq s_3$.

\section{Experimental Results}
\label{ch:Results}

The main result presented herein is the characterization of the magnitude of the third normalized Stokes parameter $|s_3|$ of a paraxial ARPB with unity relative amplitude, $\hat{V}=1$, and different values of the phase parameter $\psi$. The figures in Fig.~\ref{fig:ThirdStokesParameter} depict the predicted and experimental values of $|s_3|$ on the transverse plane for ARPBs with $\psi=-\pi/2, -\pi/4, 0, \pi/4, \pi/2$. The first row, (a)-(e), illustrates the theoretical $s_3$ of an ARPB at focus, while the second row, (f)-(j), presents its magnitude, $|s_3|$, calculated via experimental results. The colorbars on the right of the figures illustrate that blue represents a negative $s_3$ value, and red signifies a positive one. In the experimental data, the sign of $s_3$ is selected to more effectively display the variation in $s_3$ for an ARPB as $\psi$ changes.
The theoretical predictions display an annular ring around the beam center, with maximum absolute values for $\psi=\pm\pi/2$, and zero for $\psi=0$. Intermediate values are observed for ARPBs with $\psi=\pm\pi/4$. The experimental results align consistently with the theoretical predictions, particularly for $\psi=\pm\pi/2$ and $\psi=\pm\pi/4$. Although we anticipated $|s_3|$ to be null across the transverse plane for an ARPB with $\psi=0$, a residual $|s_3|$ is evident in experimental measurements, albeit at much lower intensities than those at $\psi=\pm\pi/4$. We speculate that this residue stems from the unwanted presence of small anisotropic effects within the optical setup shown in Fig.~\ref{fig:opticalsetup}. These additional phase shifts between the $x$ and $y$ field components result in non-zero contributions to the third normalized Stokes parameter $s_3$. 

\begin{figure}
    \centering
    \includegraphics[width=\linewidth]{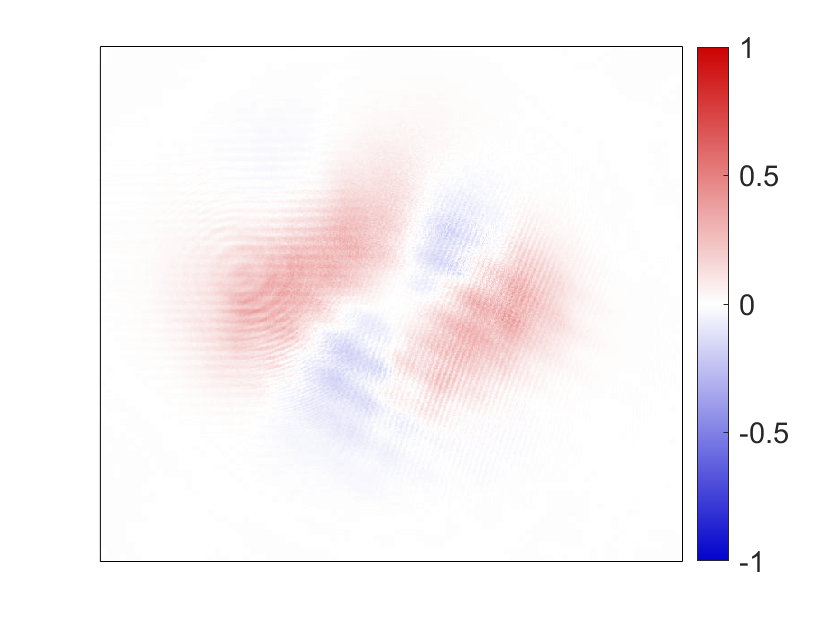}
    \caption{Second normalized Stokes parameters $s_2$ of an ARPB with $\hat{V}=1$ and $\psi=0$ after a QWP, equivalent to $s_3$ without the QWP. On average, $s_2\approx 0$, confirming the lack of chirality of ARPBs with $\psi=0$.}
    \label{fig:QWP}
\end{figure}

To confirm that the measured residue for the $\psi=0$ ARPB does not indicate a real chirality density in the transverse plane, we add a quarter-wave plate (QWP) with the fast axis on the horizontal ($x$) axis before the polarization camera in the setup from Figure~\ref{fig:opticalsetup}. The QWP transforms left/right-handed circularly polarized light into anti/diagonally polarized light. Therefore, in the new analysis, the normalized helicity density of the paraxial ARPB is proportional to the second normalized Stokes parameter $s_2$ in the imaging plane. This behavior is depicted in Figure~\ref{fig:QWP}, which displays the measured $s_2$ after a quarter-wave plate (QWP) with the fast axis at $0^{\degree}$ for an ARPB with $\psi=0$. Indeed, it is shown in Figure~\ref{fig:QWP} that $s_2\approx 0$ on average, and that the residue shown in Fig.~\ref{fig:ThirdStokesParameter}(h) is not indicative of a non-zero helicity density but rather of small anisotropic effects within the optical setup that distort the local polarization of the field.

\begin{figure}
    \centering
    \includegraphics[width=\linewidth]{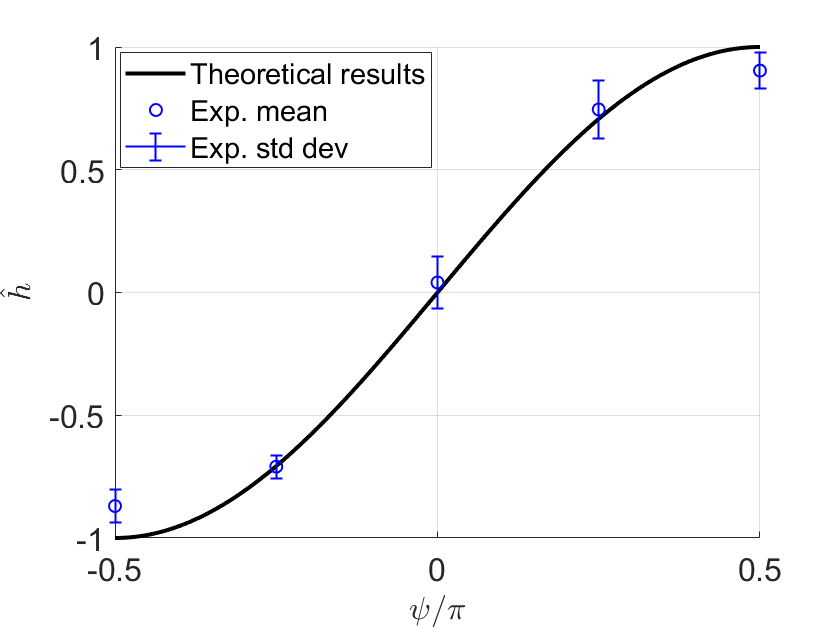}
    \caption{Normalized helicity density $\hat{h}$ plotted against phase parameter $\psi$ for an ARPB with $\hat{V}=1$. The black line shows theoretical values computed from Eq.~(\ref{eq:ARPBNormalizedHelicityDensity}), while the blue circles represent averaged experimental values, and the vertical blue lines indicate their standard deviations. The sign of the experimental $s_3$ has been included for illustrative purposes. Varying $\psi$ it is possible to obtain any value of helicity density.}
    \label{fig:sinPsi}
\end{figure}

Figure~\ref{fig:sinPsi} illustrates the relationship between the normalized helicity density $\hat{h}$ and the phase parameter $\psi$ of an ARPB with $\hat{V}=1$. The theoretical values, represented by a black line, follow the sinusoidal dependence from Eq.~(\ref{eq:ARPBNormalizedHelicityDensity}). The calculated values of $|\hat{h}|$ are obtained by normalizing the experimental $|S_3|$ by their respective $S_0$ (as shown in Fig.~\ref{fig:ThirdStokesParameter}), where the values of $S_0$ below a certain threshold are neglected to avoid division by zero outside of the beam. 

The average of the resulting values of $|\hat{h}|$ is denoted by a blue circle, with vertical blue lines indicating the standard deviations of the filtered $|\hat{h}|$ for each value of $\psi$. In the figures presented we depicted the quantity $\hat{h}$ (and equivalently $s_3$) under the assumption that we have knowledge of its sign. This assumption is made for illustrative purposes to visually represent the change in the normalized helicity density of the experimental ARPB with respect to the value of its phase parameter $\psi$.

\begin{figure}[htbp]
    \centering

    \setlength{\tabcolsep}{1pt} 

    \begin{tikzpicture}
        \node[draw=black, thick, inner sep=2] {
            \begin{tabular}{cc}
                \includegraphics[width=0.48\linewidth]{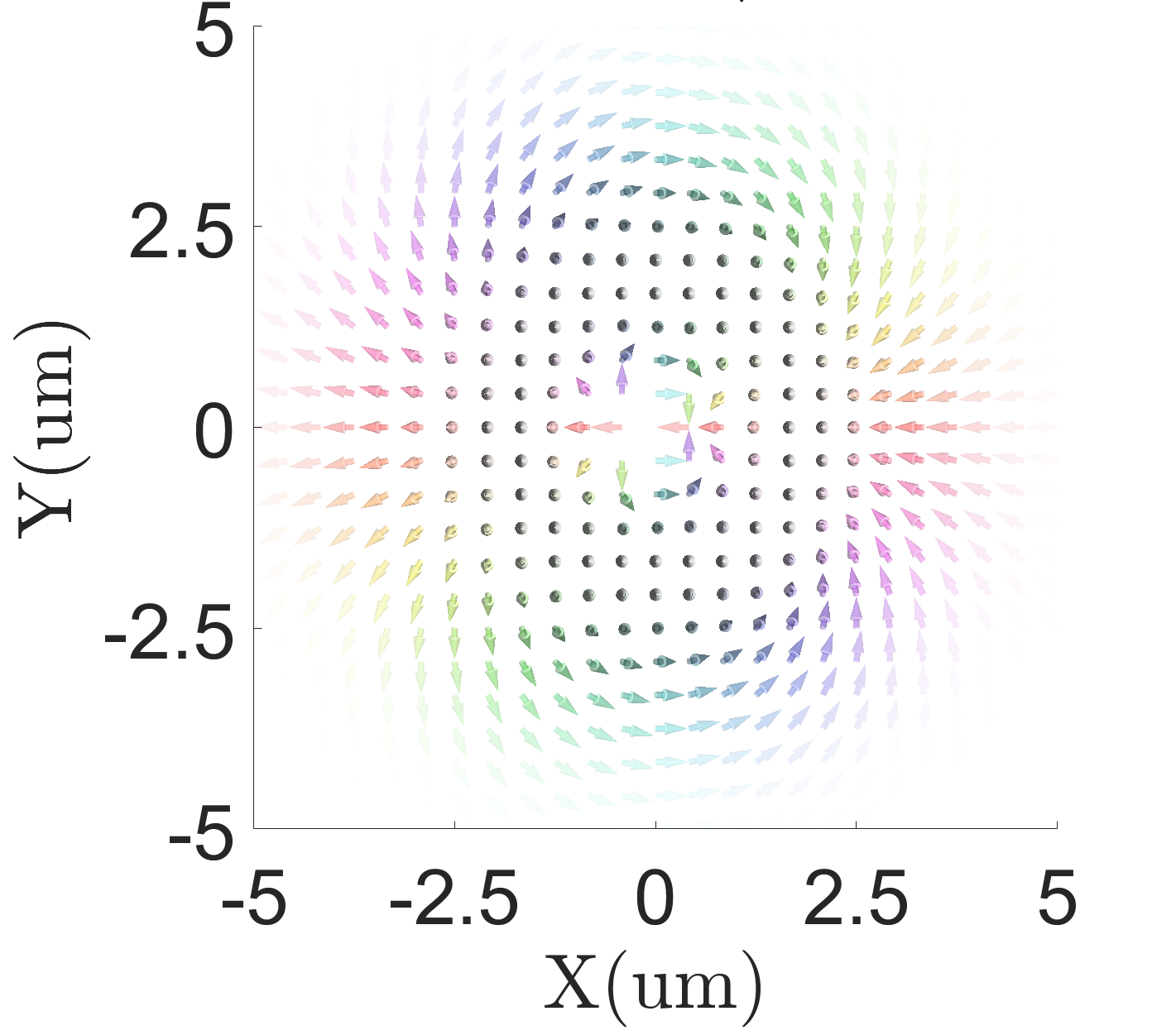} & 
                \includegraphics[width=0.48\linewidth]{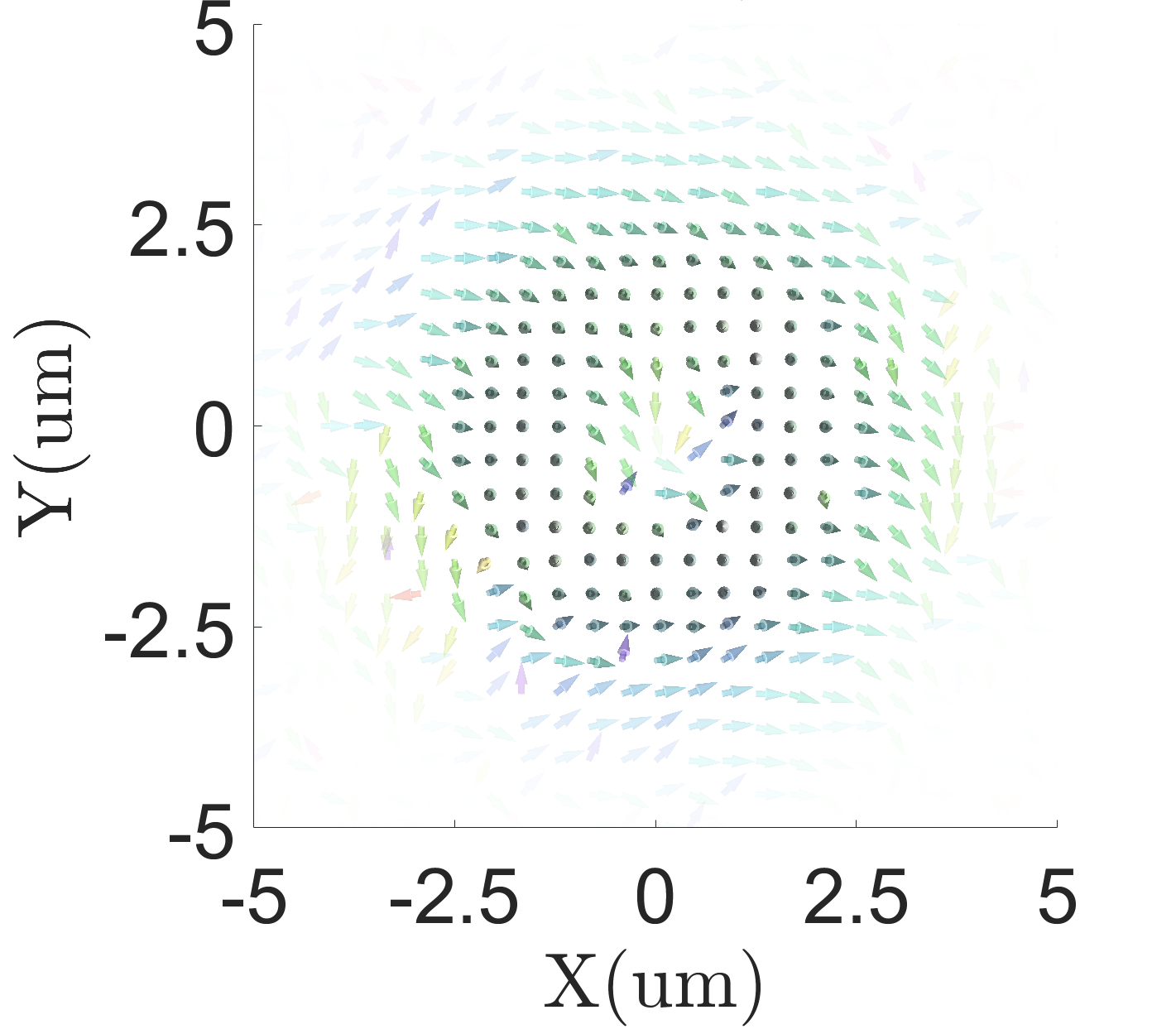} \\
                \multicolumn{2}{c}{(a) $\psi = -\frac{\pi}{2}$} \\ 
            \end{tabular}
        };
    \end{tikzpicture}

    \begin{tikzpicture}
        \node[draw=black, thick, inner sep=2] {
            \begin{tabular}{cc}
                \includegraphics[width=0.48\linewidth]{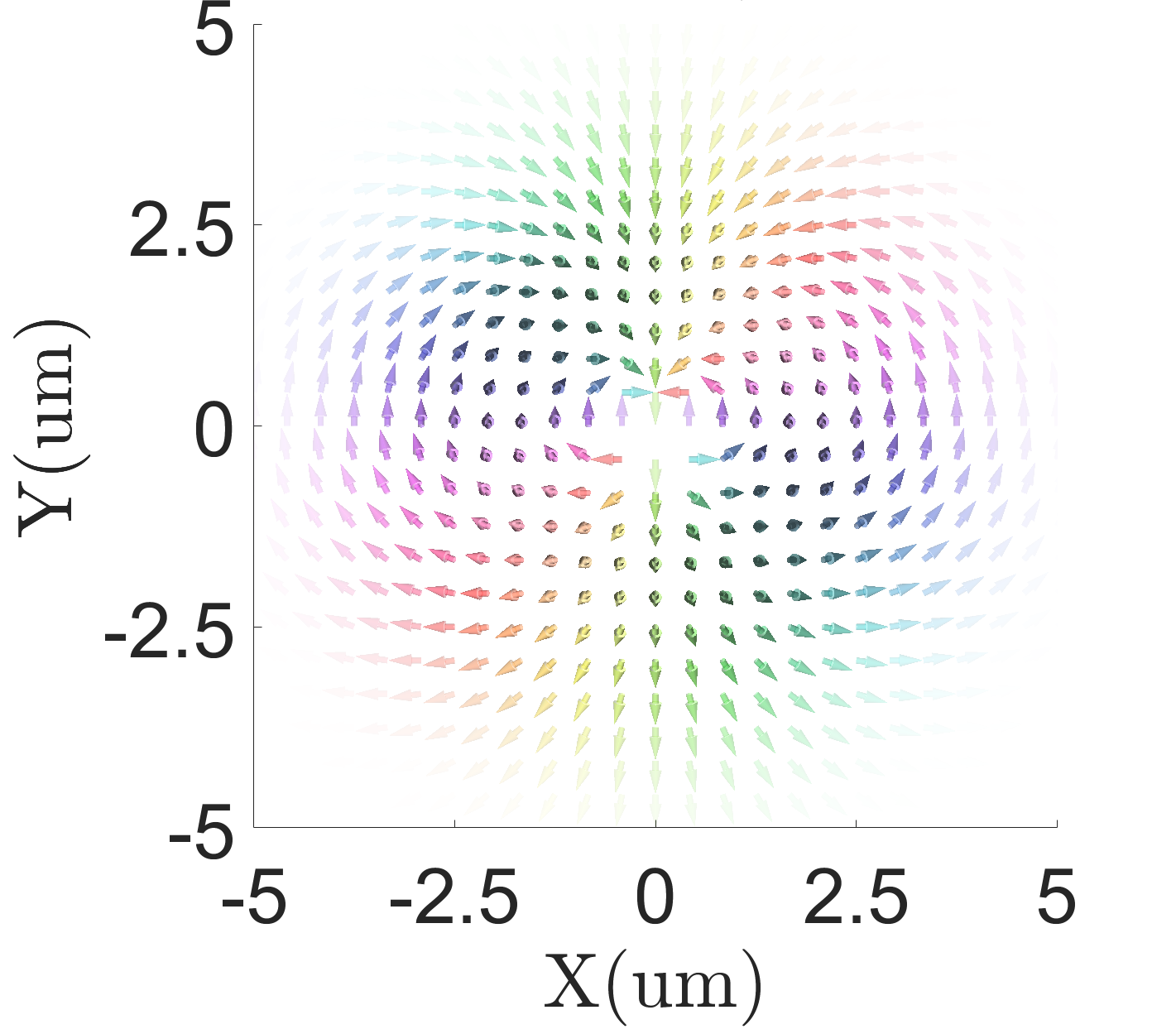} & 
                \includegraphics[width=0.48\linewidth]{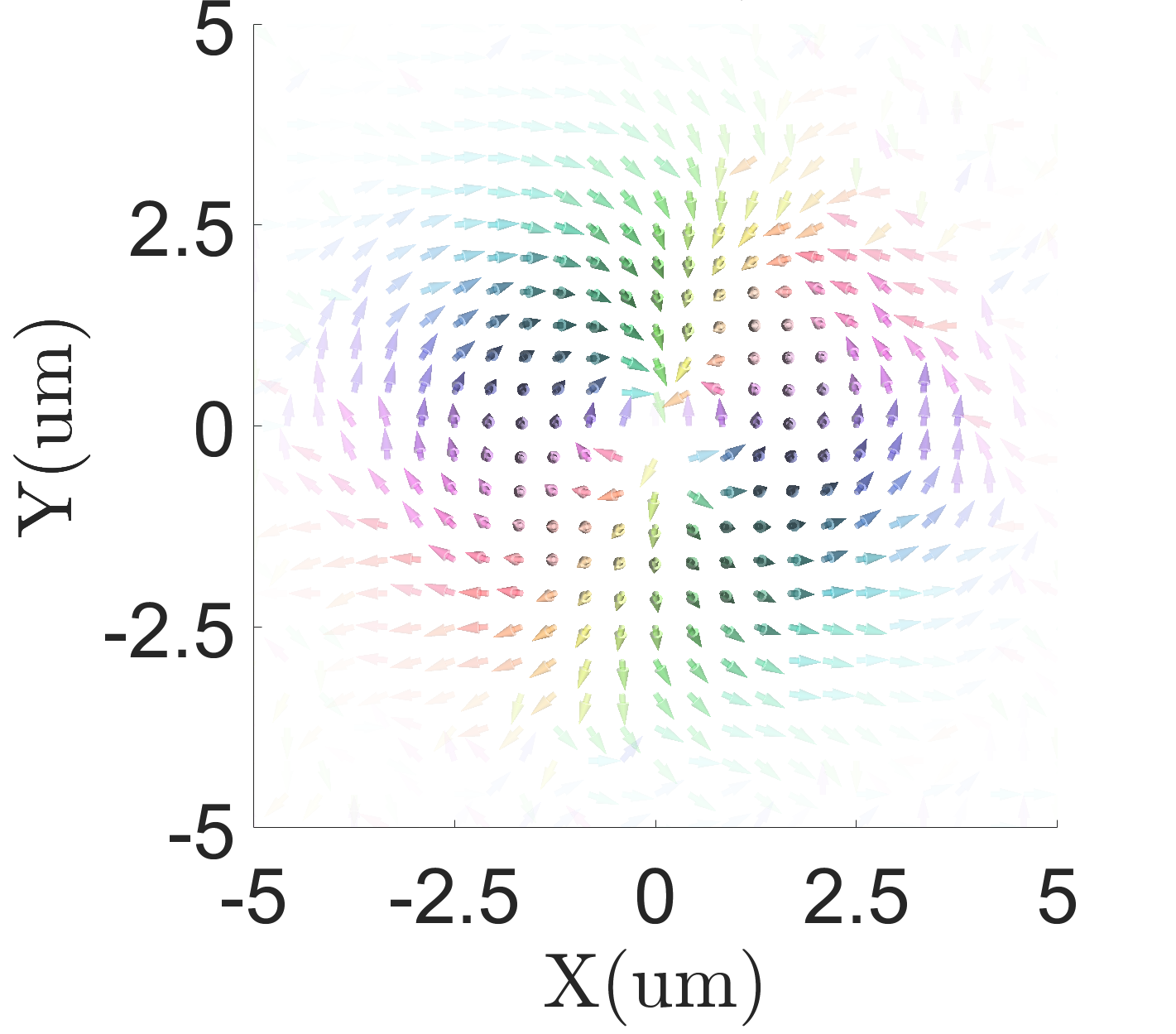} \\
                \multicolumn{2}{c}{(b) $\psi = -\frac{\pi}{4}$} \\ 
            \end{tabular}
        };
    \end{tikzpicture}

    \begin{tikzpicture}
        \node[draw=black, thick, inner sep=2] {
            \begin{tabular}{cc}
                \includegraphics[width=0.48\linewidth]{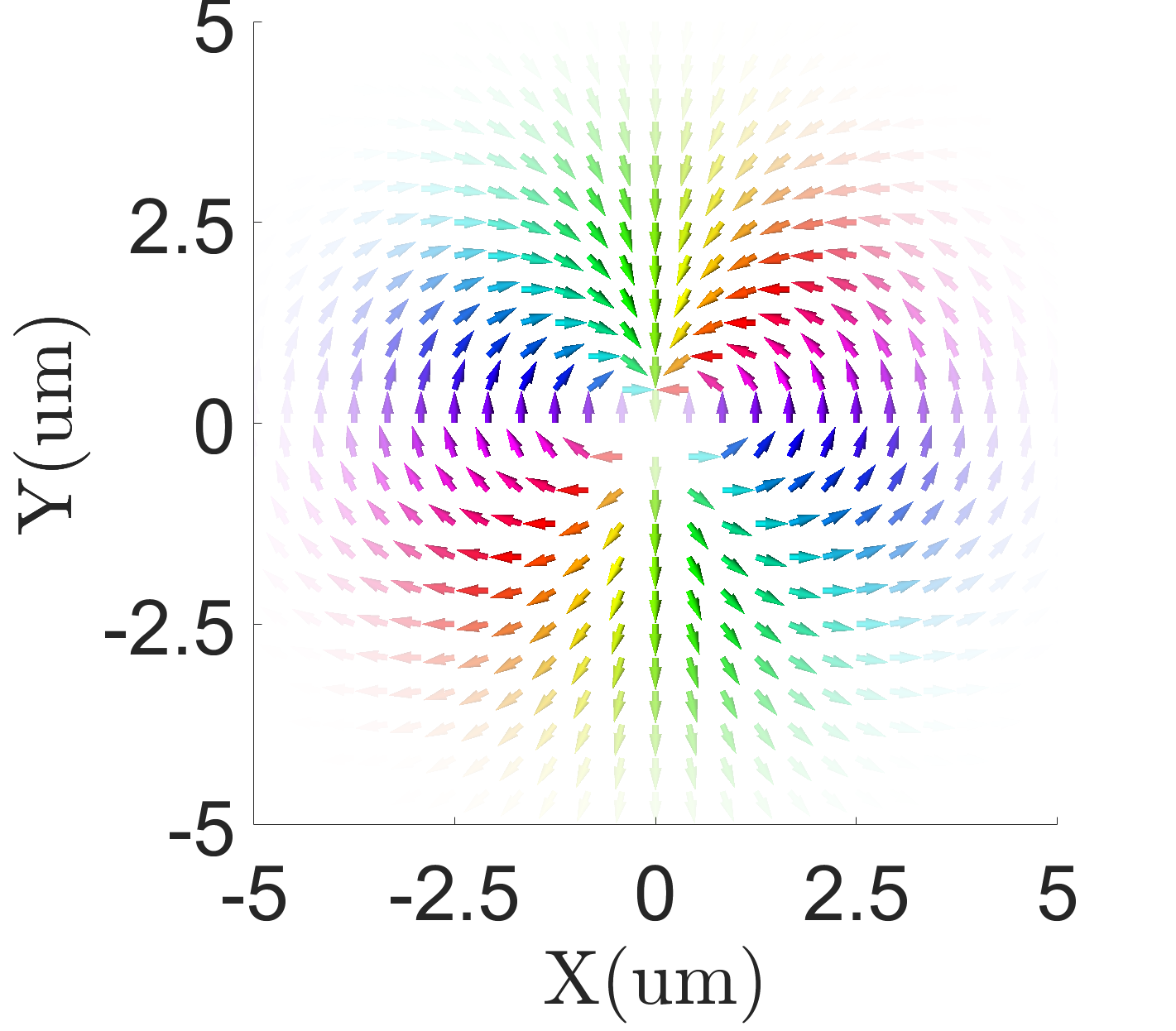} & 
                \includegraphics[width=0.48\linewidth]{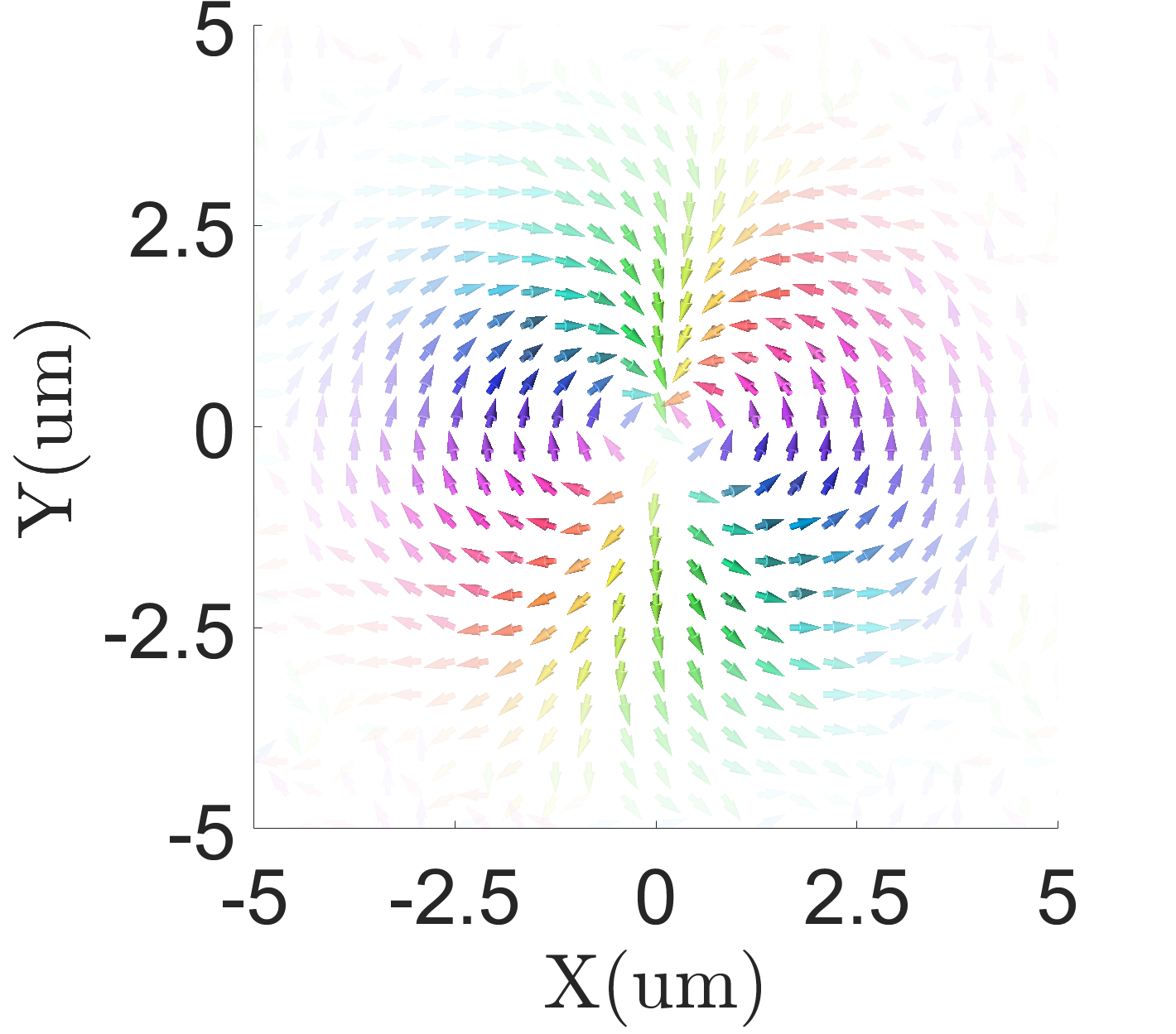} \\
                \multicolumn{2}{c}{(c) $\psi = 0$} \\ 
            \end{tabular}
        };
    \end{tikzpicture}

    \begin{tikzpicture}
        \node[draw=black, thick, inner sep=2] {
            \begin{tabular}{cc}
                \includegraphics[width=0.48\linewidth]{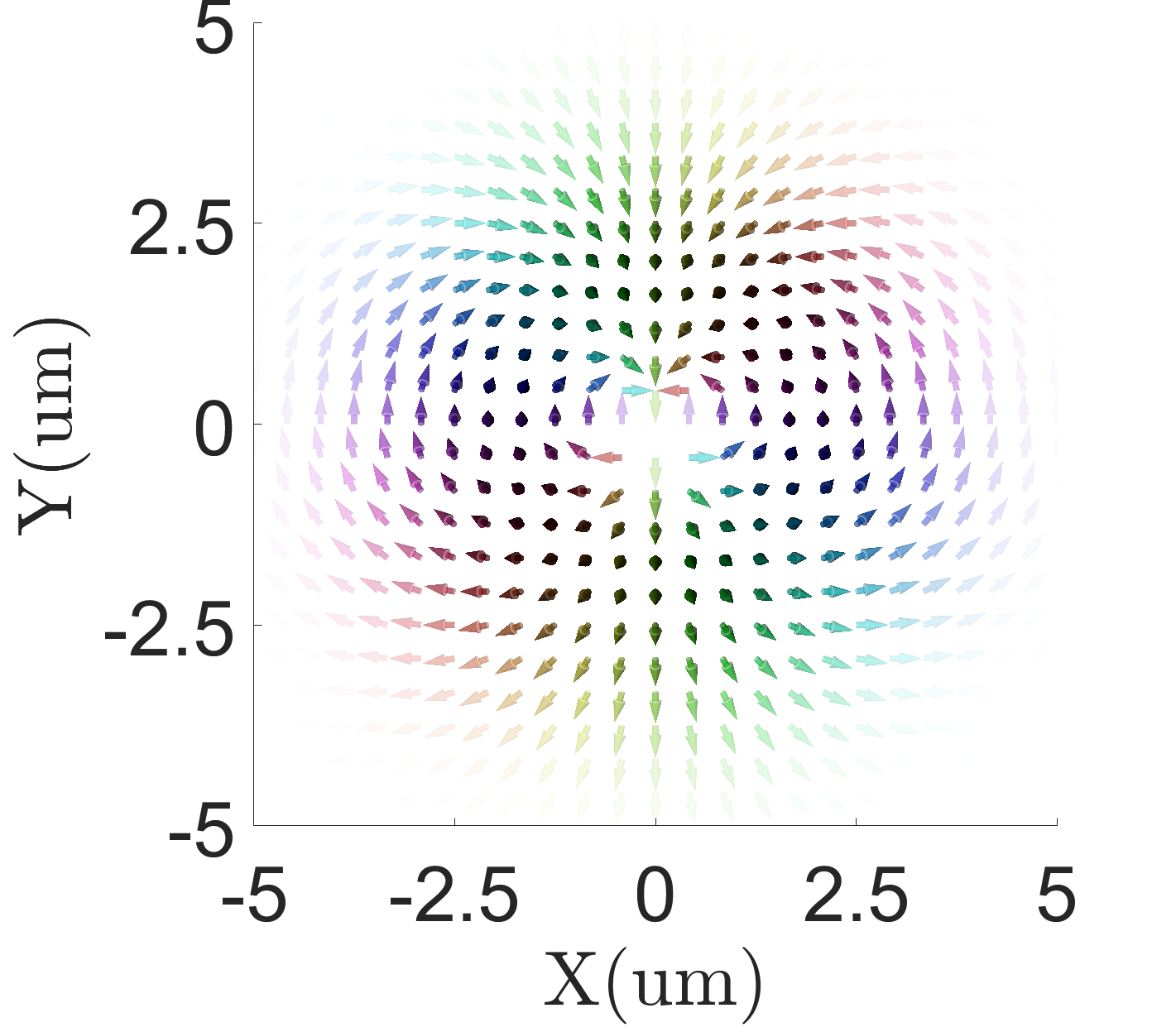} & 
                \includegraphics[width=0.48\linewidth]{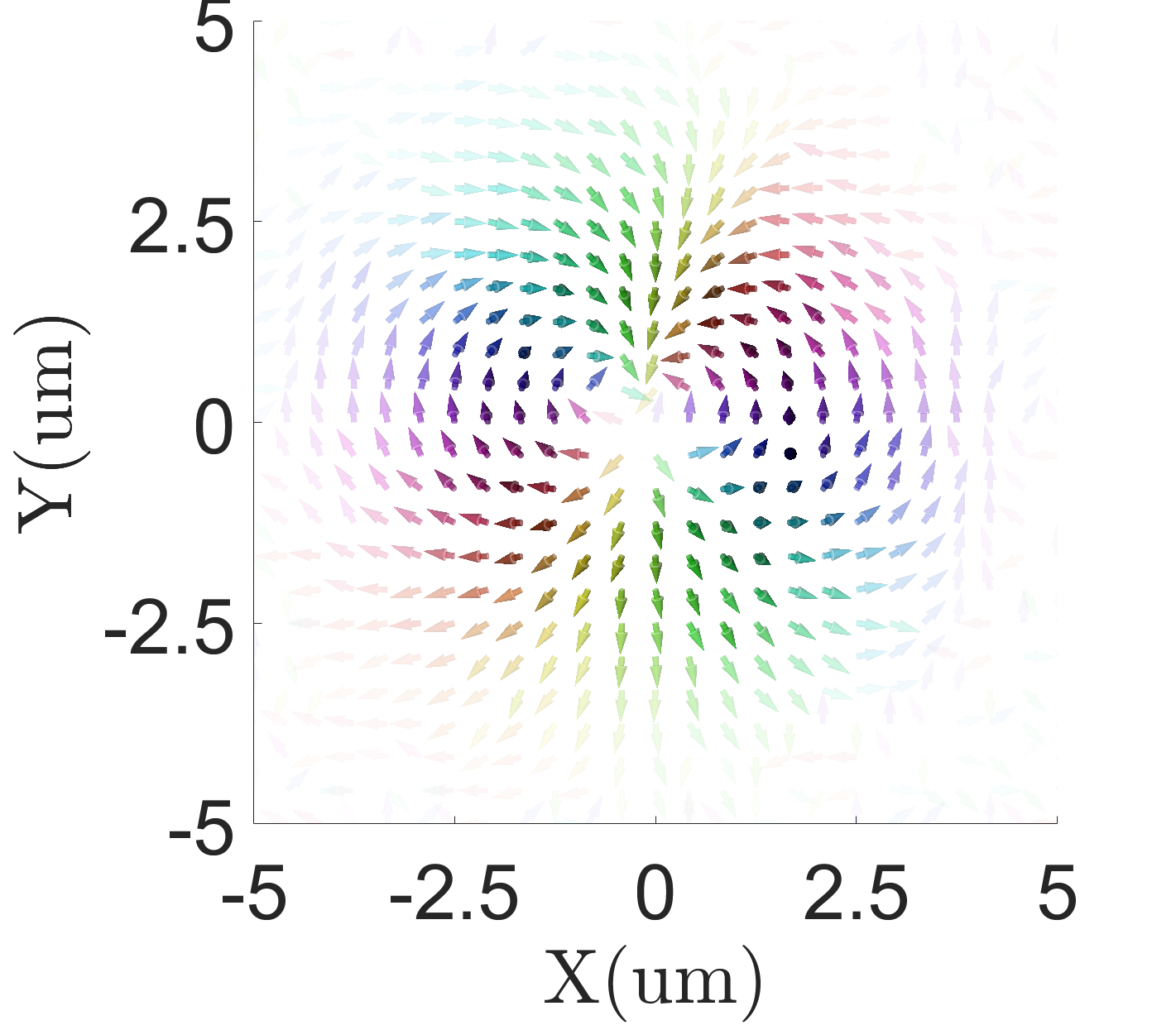} \\
                \multicolumn{2}{c}{(d) $\psi = \frac{\pi}{4}$} \\ 
            \end{tabular}
        };
    \end{tikzpicture}

    \begin{tikzpicture}
        \node[draw=black, thick, inner sep=2] {
            \begin{tabular}{cc}
                \includegraphics[width=0.48\linewidth]{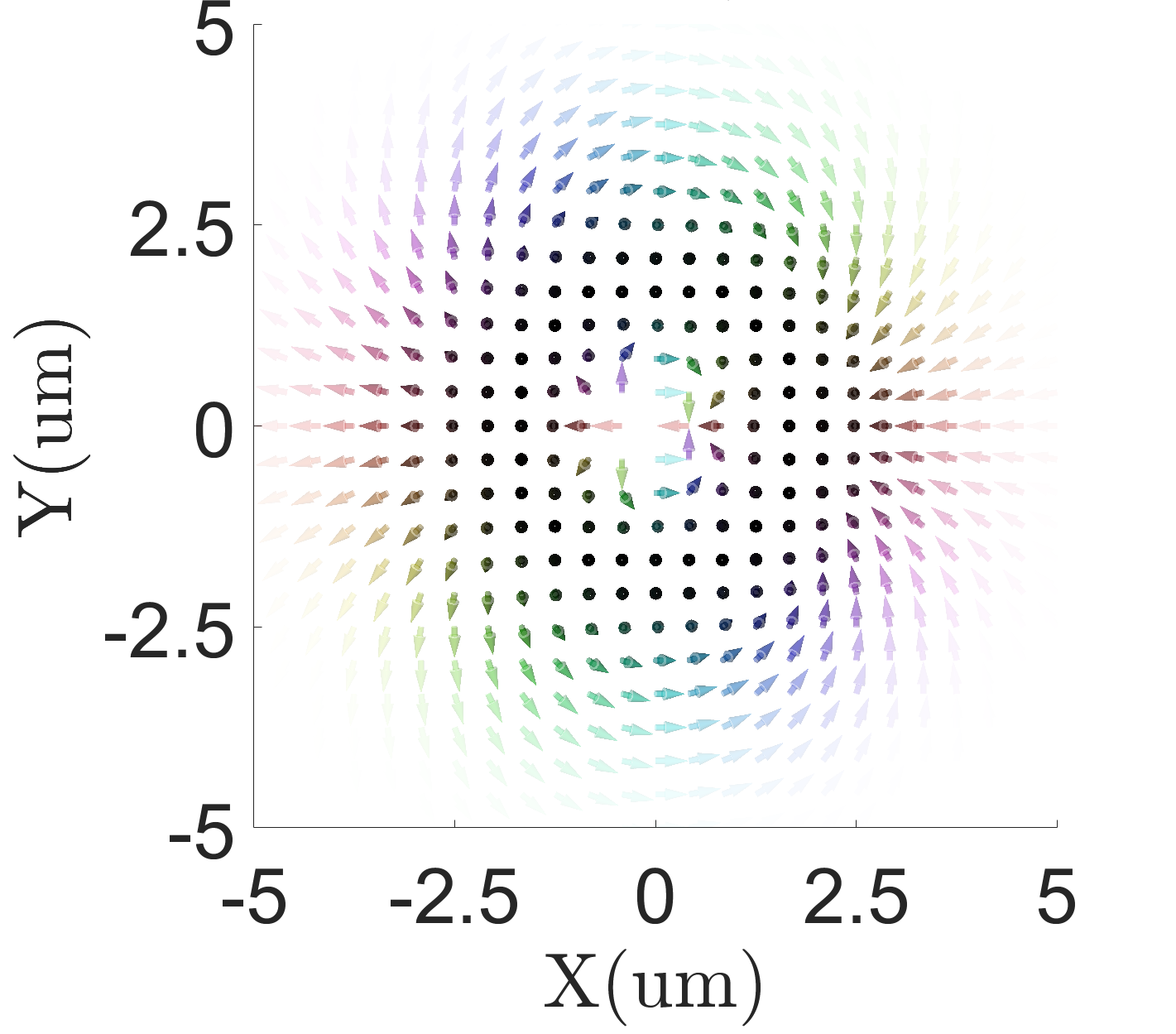} & 
                \includegraphics[width=0.48\linewidth]{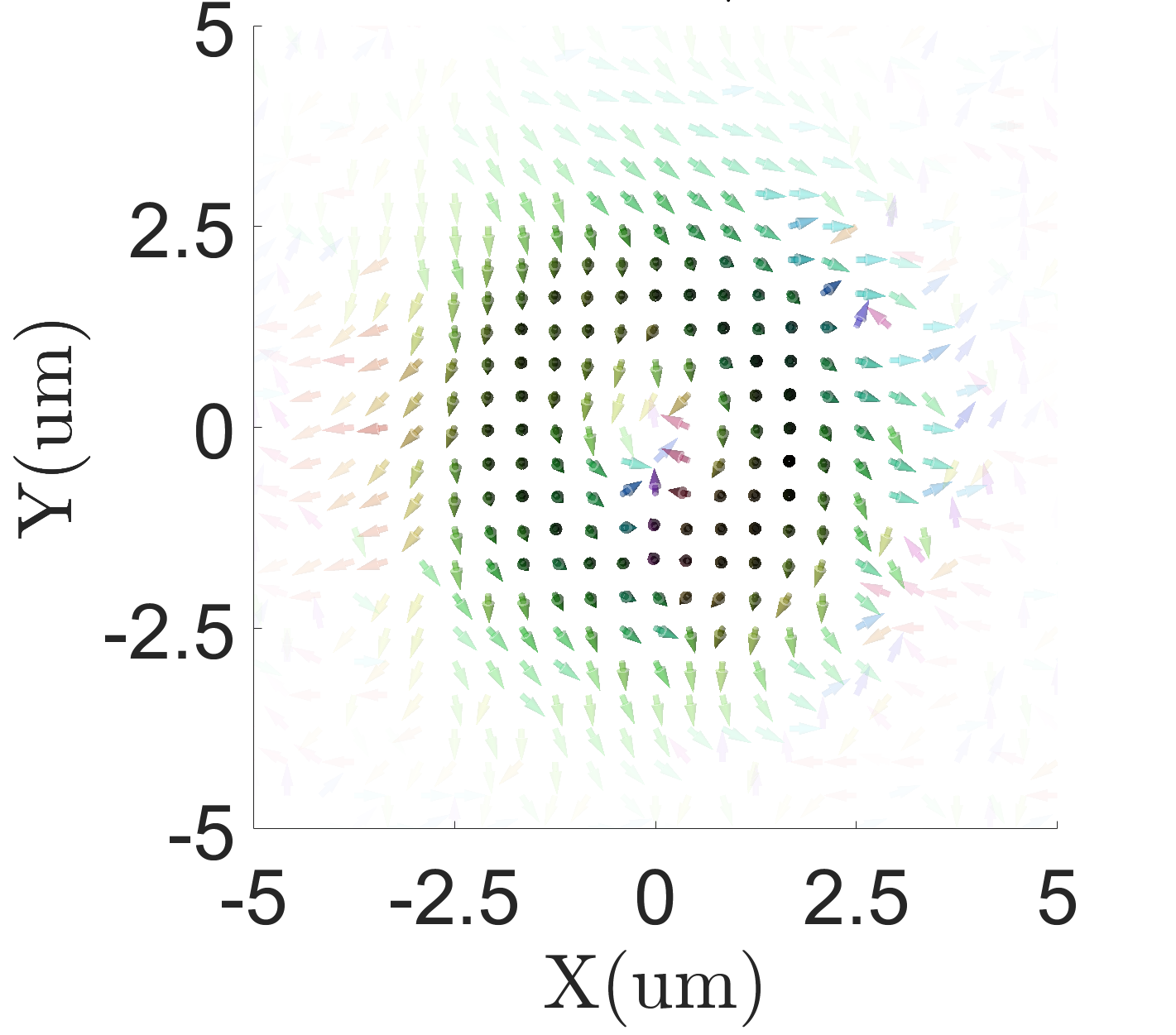} \\
                \multicolumn{2}{c}{(e) $\psi = \frac{\pi}{2}$} \\ 
            \end{tabular}
        };
    \end{tikzpicture}

    \caption{Theoretical (left) and experimental (right) polarization textures of the transverse electric fields at the focus of an ARPB varying the phase parameter $\psi$. The figures are arranged in order of increasing $\psi$ from $-\pi/2$ to $\pi/2$. Each pair is enclosed within a bordered box for clarity. Dots represent circular polarization, arrows fully in the $x,y$ plane represent linear polarization, and arrows oblique to the $x,y$ plane represent elliptical polarization.}
    \label{fig:PolText}
\end{figure}

The local polarization of structured light can be visualized using polarization textures where the angle and ellipticity of the electric field at a point are represented by projecting the arrows (normalized) pointing from the center of the Poincaré sphere to the direction of the respective polarization state on the Poincaré sphere. Linear polarization is represented as arrows fully in the $x,y$ plane, circular polarization as arrows on the $z$ axis (so only a dot is visible), and elliptical polarization is represented as arrows outside of the $x,y$ plane (projected arrows are shorter than those representing linear polarization that have full length). These arrows are then plotted at various points in the transverse $x,y$ plane of the beam, revealing the texture. The polarization textures of the implemented ARPBs are shown in Figure~\ref{fig:PolText}, with theoretical predictions on the left and experimental results on the right. The figures are organized in pairs with the same value of $\psi$ enclosed in bordered boxes, and arranged vertically in increasing order of $\psi$, ranging from $-\pi/2$ to $\pi/2$. We observe that the theoretical OC-ARPBs (with $\psi=\pm\pi/2$, shown in the left of Figure~\ref{fig:PolText}(a) and (e)) exhibit local circular polarization away from the beam edges. As the phase parameter $\psi$ approaches zero, these chiral polarizations transition to linear polarizations. The lack of circularly polarized components for an ARPB with $\psi=0$ is proof of its lack of chirality. Indeed, we see that the degree of circular polarization away from the center and  edges of the beam reaches a maximum for ARPBs with $\psi=\pm\pi/2$, and a minimum for $\psi=0$. These results reinforce the demonstration of optimally chiral structured light discussed in relation to Figures~\ref{fig:ThirdStokesParameter} and \ref{fig:sinPsi}. By comparing the right and left columns in Figure~\ref{fig:PolText}, we can appreciate that the experimental results largely agree with the theoretical predictions, particularly away from the beam edges, where the experimental polarization textures show some variability. 

\section{Relevant properties of OCL}
\label{ch:TransverseOC}

The results presented in the previous sections are important because they have demonstrated precise control over the helicity density of the ARPB by only tuning the phase parameter $\psi$. These investigations were conducted under the paraxial approximation \cite{lax_maxwell_1975}. Since the ratio $w_0/\lambda$ is significantly larger than unity, the longitudinal fields are effectively smaller than the transverse ones in most of the transverse focal plane (see Ref.~\cite{veysi_focused_2016} for details). Furthermore, the measurement setup shown in Fig.~\ref{fig:opticalsetup} only distinguishes fields polarized in the transverse plane. Consequently, the excellent agreement between the theoretical and experimental normalized helicity densities shown in Section~\ref{ch:Results} further validates the use of the paraxial approximation and the neglecting of the longitudinal fields of an ARPB with a large beam waist parameter compared to the wavelength. Here, we expand on this aspect by showing that any field (it does not need to be a beam) that satisfies the optimal chirality condition from Eq.~(\ref{eq:OptimalChiralityCondition}) in the {\it transverse} plane, i.e., $\mathbf{E}_\perp = \pm i\eta_0\mathbf{H}_\perp$, is a total OC field. To establish this, we use Maxwell's equations to derive the expressions for the longitudinal fields in terms of the transverse fields:

\begin{equation}
    \begin{array}{c}
        H_z=\frac{-i}{\omega\mu_0}(\nabla\times\mathbf{E}_\perp)\cdot\mathbf{z}, \\
        E_z=\frac{i}{\omega\varepsilon_0}(\nabla\times\mathbf{H}_\perp)\cdot\mathbf{z}.
    \end{array}
    \label{eq:ApBLong}
\end{equation}

Assuming the field exhibits optimal chirality in the transverse plane, i.e., $\mathbf{E}_\perp = \pm i\eta_0\mathbf{H}_\perp$, and substituting this relationship into Eq.~(\ref{eq:ApBLong}), it follows that $E_z = \pm i\eta_0H_z$. The longitudinal fields display the same magnitude ratio and phase shift between the electric and magnetic fields as the transverse fields. This calculation shows that optimal chirality in the transverse plane necessarily extends to the longitudinal direction as well.

For an ARPB, all field components display the same phase shift between the electric and magnetic fields regardless of whether it is optimally chiral. This is observed by following Ref.~\cite{parareda_arpb_2024}, where the helicity density was decomposed into its component contributions as $h = h_\rho + h_\varphi + h_z$, where each component is $h_n = \Im(E_nH_n^*)/(2\omega c)$, for $n = \rho, \varphi, z$. Specifically, for the ARPB we have

\begin{equation}
    \begin{array}{c}
         h_\rho = h_0(k\rho)^2\left(A_\rho^2 + B_\rho^2\right) \hat{V} \sin\psi,\\
         h_\varphi = h_0(k\rho)^2 \hat{V} \sin\psi,\\
         h_z = 4h_0\left(A_z^2 + B_z^2\right) \hat{V} \sin\psi.
    \end{array}
    \label{eq:HelicityDensityComponents}
\end{equation}

The parameters $A_\rho, A_z, B_\rho, B_z$  from Eq.~(\ref{eq:Simplification}) and $\hat{V}$ are real-valued, and therefore the sign of each helicity density component $h_n$ is solely determined by the term $\sin\psi$. This means that for an ARPB, all components $h_n$ share the same sign of helicity density, hence they all contribute constructively to having a specific helicity sign.

Additionally, optimally chiral structured light can be obtained without circular polarization, as it happens for a focused OC-ARPB  \cite{parareda_arpb_2024}. Therefore, circular polarization is not a necessary condition for three-dimensional light to be optimally chiral.  

Here, we further demonstrate that an optimally chiral beam under the paraxial approximation is necessarily circularly polarized. This latter statement is demonstrated by combining the optimal chirality condition, $\mathbf{E}=\pm i \eta_0\mathbf{H}$, with the Faraday equation, leading to 

\begin{equation}
    \nabla\times\mathbf{E} = \pm k\mathbf{E},
    \label{eq:NablaEkE}
\end{equation}

which is a necessary and sufficient condition to having optimal chiral light \cite{parareda_arpb_2024}. Eq.~(\ref{eq:NablaEkE}) shows that optimally chiral light has fields that are eigenvectors of the curl operator \cite{lekner_self_dual_2024}. Therefore, monochromatic optimally chiral beams are self-dual \cite{lekner_selfdual_2019}.

For a three-dimensional field, using  cartesian coordinates, i.e., $\mathbf{E} = E_x\hat{\mathbf{x}} + E_y\hat{\mathbf{y}} + E_z\hat{\mathbf{z}}$, we have

\begin{equation}
    \begin{array}{c}
       \frac{\partial E_z}{\partial y} - \frac{\partial E_y}{\partial z} = \pm k E_x, \\
       \frac{\partial E_x}{\partial z} - \frac{\partial E_z}{\partial x} = \pm k E_y, \\
       \frac{\partial E_y}{\partial x} - \frac{\partial E_x}{\partial y} = \pm k E_z. 
    \end{array}
    \label{eq:curlofE}
\end{equation}


Under the paraxial approximation, $\frac{\partial }{\partial z}>> \frac{\partial}{\partial y} , \frac{\partial}{\partial y}$ \cite{jisha_paraxial_2017} (where the authors show the phase variation along an axis in terms of the wavenumber), and assuming that the field propagates along $z$ as $\mathbf{E}\propto e^{ikz}$, the equations in Eq.~(\ref{eq:curlofE}) are simplified to

\begin{equation}
    \begin{array}{c}
        -ikE_y = \pm kE_x, \\
        ikE_x = \pm k E_y, \\
        |E_z|/|\mathbf{E}_\perp| <<1,
    \end{array}
\end{equation}

implying that the field is circularly polarized, i.e., $E_y = \pm i E_x$, wherein the two transverse components have equal magnitudes and a $\pi/2$ phase delay between them. Analogous proof holds for the transverse magnetic field that is also circularly polarized. For paraxial fields whose longitudinal fields are negligible compared to the transverse field components (for intermediate radial distance from the beam axis), the concept of optimally chiral light is equivalent to that of circular polarization on the focal plane, and only for intermediate radial distance as discussed for the ARPB in the next section. However, upon the introduction of structured light with considerable longitudinal fields, optimal chirality can be obtained without circular polarization.

\section{Discussion}
\label{ch:Discussion}

The presented results not only provide the first experimental analysis and confirmation of the realizability of an optimally chiral structured beam, the OC-ARPB, but also contribute to the limited experimental investigations of self-dual fields. While the primary focus of this paper is not on self-duality, our work aligns with observations in Ref.~\cite{lekner_self_dual_2024}, where monochromatic self-dual electromagnetic fields are identified as eigenvectors of the curl operator.

The experiments described in this work demonstrate precise control over the helicity density of the ARPB via the tuning of the phase parameter $\psi$, following the relation $\hat{h} = \sin\psi$ when $\hat{V}=1$. Therefore, we have shown that the helicity density of the ARPB can be tuned across its full range of possible values, namely, $-u/\omega \leq h \leq u/\omega$, by only varying the single beam parameter $\psi$ (with $\hat{V}=1$). Modifying other beam parameters, such as $w_0$ and $\lambda$, shapes the topology of the beam instead. For example, reducing the $w_0/\lambda$ ratio leads to higher energy and helicity densities at the beam focus ($z=0$), as shown in Ref.~\cite{parareda_arpb_2024}.

Additionally, we have been able to verify the local polarization of the paraxial ARPBs on the transverse plane with the polarization textures shown in Figure~\ref{fig:PolText}. The paraxial OC-ARPBs (i.e., when $\psi = \pm \pi/2$ and $\hat{V}=1$) exhibit the highest degree of local circular polarization. The experimental results are in agreement with what was discussed after Eq.~(20) in Ref.~\cite{parareda_arpb_2024} that paraxial OC-ARPBs are primarily circularly polarized at the beam focal plane $z=0$. The transverse fields of an OC-ARPB are 

\begin{equation}
    \begin{array}{c}
         \mathbf{E}_\perp = \frac{\rho }{w^2}f V \left[\left(A_\rho + i B_\rho\right)\,\hat{\bm{\rho}} \pm i\,\hat{\bm{\varphi}}\right], \\
         \mathbf{H}_\perp = \mp i \mathbf{E}_\perp / \eta_0. \\
    \end{array}
    \label{eq:OCARPBTransverse}
\end{equation}

Here, we further observe from Eq.~(\ref{eq:Simplification}) that for OC-ARPBs with a large waist relative to the wavelength ($w_0 \gg \lambda$) on the focal plane $z=0$, one has $A_\rho\approx 1$ and $B_\rho=0$, resulting in circular polarization in the transverse plane. Note that however when $\rho \gg w_0$, then $A_{\rho}$ is not close to unity anymore. Therefore we do not have circular polarization at the edges of the beam, as shown in Fig.~\ref{fig:PolText} in both theory and experiment. 

Additionally, the transverse circular polarization of paraxial OC-ARPB is also lost near the beam axis. This result is appreciated in Figure~\ref{fig:PolText}(a), where the local polarization of the theoretical OC-ARPB becomes elliptical near the beam axis. When $\rho \ll w_0$, the longitudinal fields cannot be neglected. This occurs when the term $A_z$ is no longer much less than $k\rho A_{\rho}$ at $z=0$. Consequently, near the beam axis $\hat{h}\neq s_3$, in agreement with the property of structured OC fields displaying optimal chirality ($|\hat{h}|=1$) without being circularly polarized ($|s_3|=1$). This result is consistent with the discussion in Ref.~\cite{lekner_selfdual_2019} stating that transversely finite beams which are circularly polarized everywhere in a fixed plane do not exist.
 
In summary, while circular polarization is a sufficient condition to attain optimal chirality, it is not a necessary one since a {\em focused} OC-ARPB displays optimal chirality without circular polarization. Even when the OC-ARPB has a large beam waist ($w_0 \gg \lambda$), it still displays optimal chirality without circular polarization near the axis and far away from it.

Our investigation has directly verified the optimal chirality of the paraxial OC-ARPB with $\hat{V} = 1$ and $\psi = \pm\pi/2$ only in the transverse plane, and at distances that are not near the axis nor far away from it. Even though our setup, illustrated in Figure~\ref{fig:opticalsetup}, cannot differentiate the longitudinal components of light, the OC features of $E_z$ and $H_z$ have been theoretically demonstrated by using Maxwell's equations. Since the ARPB offers promising avenues for precise probing and manipulation of chiral particles, future research will explore the optical chirality of non-paraxial ARPBs that display strong longitudinal fields on the beam axis.

\section{Conclusion}
\label{ch:Conclusion}

We have successfully generated an ARPB using a versatile optical setup with two SLMs employing orthogonal polarizations ($x$ and $y$). By adjusting the phase parameter $\psi$, we demonstrated the ability to manipulate the chirality density of the ARPB across its full range of possible values. Notably, we found that the paraxial ARPB can achieve optimal chirality for $\psi=\pm\pi/2$, showcasing the existence of optimally chiral structured light. 

While the experiments realized herein are restricted to the transverse plane, we have also theoretically shown that three-dimensional fields whose transverse components satisfy the optimal chirality condition are optimally chiral in all directions. Additionally, we have demonstrated that circular polarization is a sufficient but not necessary condition for structured fields to be optimally chiral, and that it is equivalent to the concept of optimal chirality only under the paraxial approximation and when the longitudinal fields are negligible compared to the transverse field components. We found that the local polarization of the OC-ARPB is circular away from the center or edges of the beam. In those regions, the local circular polarization is lost even though optimal chirality is maintained.

Additionally, we have shown that monochromatic optimally chiral fields are self-dual since their electric and magnetic fields are the eigenvectors of the curl operator, leading to maximal chirality density among other self-dual electromagnetic features. The OC-ARPB generated in this work represents an example of a structured self-dual monochromatic beam, of which few have been studied experimentally.

Importantly, the results of this study verify the first practical implementation of an OC structured beam, of which the OC-ARPB is only a specific example. This new tool provides unprecedented control over fundamental chiral light-matter interactions, with future applications of enhanced sensing and manipulation of chiral particles. Given the ubiquity and importance of chirality in biology, the development of precise tools to characterize and control chiral molecules is of supreme importance in the field of biophotonics and single-isomer drug discovery. The ability to dynamically control the helicity density of the ARPB allows for the innovative design of dynamic, enantioselective optical traps. Considering the importance of the polarization of the chiral field components on light-matter interactions, future research might explore optical chirality of non-paraxial ARPBs on the beam axis, which is solely attributed to the chiral longitudinal fields.


\begin{acknowledgments}
 We would acknowledge fruitful discussions with professor Federico Capasso, and we are grateful for his contributions that inspired our work on optimal chirality as a means to maximize chirality-discriminating forces. We would like to thank the Beckman Laser Institute Foundation and the Department of Defense (DOD Grant No. 225135) for funding this research.
\end{acknowledgments}

\section*{Author Contributions}
\textbf{Albert Herrero-Parareda}: Conceptualization, Methodology, Software, Formal Analysis, Investigation, Writing - Original draft preparation.\\  
\textbf{Nicolas Perez}: Methodology, Software, Validation, Investigation, Resources, Writing - Original draft preparation.\\
\textbf{Filippo Capolino}: Conceptualization, Writing - Review and editing, Supervision.||
\textbf{Daryl Preece}: Conceptualization, Methodology, Resources, Writing - Review and editing, Supervision, and Project Administration.

\section*{Conflict of Interest}
The authors declare no conflicts of interest.\\

\section*{Data Availability}
The data underlying the results presented in this paper are not publicly available at this time but may be obtained from the authors upon reasonable request.

\bibliographystyle{ieeetr}  
\bibliography{references}   

\appendix

\section*{Supporting information}

\renewcommand{\theequation}{A.\arabic{equation}}
\renewcommand{\thefigure}{A.\arabic{figure}}
\setcounter{equation}{0}
\setcounter{figure}{0}

\subsection*{Section A: Proof of equivalency between the normalized helicity density and the degree of circular polarization for paraxial beams\label{app1.1a}}

Under the paraxial approximation, we neglect the small longitudinal $z$ components of the electric and magnetic fields \cite{lax_maxwell_1975, bliokh_transverse_2015}, i.e., $\mathbf{E}\approx \mathbf{E}_\perp$ and $\mathbf{H}\approx \mathbf{H}_\perp$. The transverse electromagnetic fields of a generic paraxial beam are of the form

\begin{equation}
    \begin{array}{c}
        \mathbf{E}_\perp = A(\rho,\varphi, z)\left(\frac{\mathbf{x} + m\mathbf{y}}{\sqrt{1+|m|^2}}\right),  \\
        \mathbf{H}_\perp \approx \left(\mathbf{z}\times\mathbf{E}_\perp\right)/\eta_0,
    \end{array}
    \label{eq:ParaxialBeam}
\end{equation}

where $m = E_y/E_x$ and $A(\rho,\varphi,z) = E_x\sqrt{1+|m|^2}$ are complex valued. The time-average energy density $u$ is \cite{angelsky_structured_2020}

\begin{equation}
    u = \frac{\varepsilon_0}{4}|\mathbf{E}|^2 + \frac{\mu_0}{4}|\mathbf{H}|^2,
    \label{eq:EnergyDensityDefinition}
\end{equation}

and the time-average helicity density is \cite{hanifeh_optimally_2020}

\begin{equation}
    h = \frac{1}{2\omega c}\Im\left(\mathbf{E}\cdot\mathbf{H}^*\right).
    \label{eq:HelicityDensityDefinition}
\end{equation}

Substituting the fields from Eq.~(\ref{eq:ParaxialBeam}) into Eqs.~(\ref{eq:EnergyDensityDefinition}) and (\ref{eq:HelicityDensityDefinition}), we obtain

\begin{equation}
    \begin{array}{c}
        u = \frac{\varepsilon_0}{2}|A(\rho,z)|^2, \\
        h = \frac{1}{2\omega c}\frac{|A(\rho,z)|^2}{\eta_0}\frac{2\Im(m)}{\sqrt{1+|m|^2}}. \\ 
    \end{array}
    \label{eq:EnergyHelicityParaxialBeam}
\end{equation}

Note that this last expression is not a good approximation for tightly focused beams where the longitudinal components are not negligible. In terms of the fields from Eq.~(\ref{eq:ParaxialBeam}), the normalized Stokes parameters are \cite{bliokh_extraordinary_2014, hayat_lateral_2015}

\begin{equation}
    \begin{array}{c}
        s_1 = \frac{1-|m|^2}{1+|m|^2}, \\
        s_2 = \frac{2\Re(m)}{1+|m|^2}, \\
        s_3 = \frac{2\Im(m)}{1+|m|^2}. 
    \end{array}
    \label{eq:NormStokesDefinition}
\end{equation}

For monochromatic beams, $s_1^2 + s_2^2 + s_3^2 = 1$. These three parameters describe the degree of the $x/y$ linear polarizations, 45º/-45º linear polarizations, and left-hand/right-hand circular polarizations, respectively \cite{bliokh_transverse_2015}.  Using the concept of  normalized helicity density $\hat{h}=h\omega/u$, introduced in Ref.~\cite{parareda_arpb_2024}, we find that for paraxial beams with negligible longitudinal fields, $\hat{h}=s_3$. For non-paraxial beams, $\hat{h}\neq s_3$, since $S_3$ does not consider the longitudinal fields, which do contribute to the helicity density. This is especially the case for focused ARPBs with waist dimensions comparable to the wavelength.

\end{document}